\documentclass[5p]{elsarticle}

\usepackage{prletters}
\usepackage{amsfonts,amsmath,amssymb,amsthm}
\usepackage{bm}
\usepackage{url}
\usepackage{comment}

\usepackage{enumerate}
\usepackage[utf8]{inputenc}
\usepackage{mathtools}
\usepackage{multirow,tabularx}
\usepackage{paralist}
\usepackage{booktabs}
\usepackage[disable]{todonotes}

\newif\ifplnote

\plnotefalse

\definecolor{Set1g}{HTML}{1b9e77}
\definecolor{Set1b}{HTML}{377eb8}
\definecolor{Set1c}{HTML}{277e77}

\usepackage[
	colorlinks=true
	,breaklinks
	]{hyperref}

\begin{document}

\title{RiskNet: Neural Risk Assessment in Networks of Unreliable Resources}

\author[AGH]{Krzysztof Rusek\corref{cor1}}
\cortext[cor1]{Corresponding author.}
\ead{krusek@agh.edu.pl}

\author[AGH]{Piotr Boryło}
\ead{piotr.borylo@agh.edu.pl}
\author[AGH]{Piotr Jaglarz}
\ead{pjaglarz@agh.edu.pl}
\address[AGH]{AGH University of Science and Technology, Institute of Telecommunications, Al. Mickiewicza 30, 30-059 Kraków, Poland;}
\author[TUM]{Fabien Geyer}
\ead{fgeyer@net.in.tum.de}
\address[TUM]{Technical University of Munich, Department of Informatics, Boltzmannstr. 3, 85748 Garching bei München, Germany;}
\author[UPC]{Albert Cabellos}
\ead{alberto.cabellos@upc.edu}
\address[UPC]{Barcelona Neural Networking, Universitat Politècnica de Catalunya, Barcelona, Spain}
\author[AGH]{Piotr Chołda}
\ead{piotr.cholda@agh.edu.pl}


\begin{abstract}
	We propose a graph neural network (GNN)-based method 
    to predict the distribution of penalties 
	induced by outages in communication networks, where connections are protected by resources shared between working and backup paths.  
	
	The GNN-based algorithm is trained only with  random graphs generated with the Barab\'{a}si--Albert model. Even though, the 
	obtained test results show that we can precisely model the penalties in a wide range of various existing topologies. 
	
    GNNs eliminate the need to simulate complex outage scenarios for the network topologies under study.
    In practice, the whole design operation is limited by 4\,ms on modern hardware. 
    This way, we can gain as much as over 12\,000 times in the speed improvement.

\end{abstract}

\begin{keyword}
	Graph neural networks (GNNs) 
    \sep message-passing neural network \sep resilience management \sep risk engineering \sep shared protection.
\end{keyword}

\tnotetext[t1]{Funding: This work was supported by the Polish Ministry of Science and Higher Education with the subvention funds of the Faculty of Computer Science, Electronics and Telecommunications of AGH University and by the PL-Grid Infrastructure.}

\maketitle

\markboth{TODO}%
{Rusek \MakeLowercase{\textit{et al.}}: RiskNet: Neural Risk Assessment in Networks of Unreliable Resources}

\section{Introduction} \label{sec:introduction}

Applications of machine learning (ML) in the area of communication networks\footnote{To avoid equivocation, we will denote `communication network' as the `topology'. The word `network' will be limited to `neural network' only.} (IP/MPLS-based, optical transport networks and the like) focus mainly on enabling data-driven self-management. The aim is to enable a network infrastructure to intelligently react in the presence of random and adversary events. We follow this path and contribute 
with a concept of \textit{resilience-aware network design}, where randomness relates to failures and consecutive recovery events. 
Here, we elaborate on provisioning  an efficient method to predict the quality of resilience when complex recovery settings are provided.

In communication networks, it is typically assumed that resilience to outages is based on the so-called protection, where each connection uses a precalculated pair of working and backup paths. The former is used before the outage and the latter serves afterwards to bypass faulty components (routers, cross-connects, links, etc.). Resilience provisioning can be based on dedicated protection, where backup resources are designated to be used in the case of faults in working paths and there is no danger of their shortage.
However, this method is extremely costly in term of infrastructure. Furthermore, the prediction of its behaviour is easy to model (we can assume independence between connections). On the contrary, a practical method, yet complex in modelling, is based on the so-called \emph{shared backup path protection} (SBPP). 
The notion of `sharing' consists in having the same backup resource pool for various connections, this makes the connections dependent.

From the application viewpoint, i.e., communication network management and operations, this approach can incur the penalties imposed on a network operator due to outages. They happen when a backup path cannot be established for a connection in the presence of resource shortage. Typically, a penalty is a monetary value proportional to the outage time experienced by a user due to a given service level agreement (SLA). 
From a business point of view, this is a random expense that should be included in the risk management process.

In this paper, the penalties are treated as the main performance metric to assess the quality of resilience and its financial aspect. 
The used modeling has to take into account probabilistic estimates (similar to the traditionally used availability, mean downtime, etc.), but also considers the amplitude of the outage impact in order to use an indicator inspired by \emph{risk engineering}. 
Since SBPP has no known analytical results for those indicators (e.g. expected total downtime per year), the only solution is to use time-consuming simulations. Due to the rare nature of failures, the simulation must be very long as according to the best authors' knowledge, there is no rare event simulation technique addressing SBPP.

The main contribution of this paper is proposition to pose the risk estimation as a regression problem defined on a bipartite meta-graph solved with the help of a graph neural network (GNN).
The GNN maps reliability parameters of network's components (e.g. links) to parameters of a distribution family.
Such a distribution is a universal tool to determine various performance characteristics for any network topology. 
By `universal', we mean that a single instance of the trained GNN can be effectively used with various network topologies, no matter what kind of topologies were used during the training process. This way, a high level of generalization is obtained, increasing the applicability potential of the proposed approach. 
We have also eliminated the need for long simulations as they are needed only during the training phase.
Inference is extremely fast as since the GNN outputs the entire distribution, we are not restricted to any particular risk measure- this is left to the decision of risk analyst.

The next section presents the literature review to show the background and emphasize the originality of our approach. Section~\ref{sec:methods} theoretically elaborates on our GNN-based approach to the prediction of penalty levels. %
Next, we devote Section~\ref{sec: evaluation-of-the-accuracy} 
for illustration of the whole experimental setup and for the presentation of the results and discussion, proving that our approach meets the practical requirements. We conclude the paper and present the ideas for the extension of the given concept in Section~\ref{sec:conclusions}.

\section{Rationale and Related Work} \label{sec:related-work}

The rationale for this work comes from the field of communication and computer networks, where data transmission must be protected against the potential impact of component failures. An operator that cannot provide reliable transmission for its customers must pay fines according to the particular service level agreements. The fee incurred in relation to a failure impact is calculated according to the so-called compensation policy. Usually, it is based on the total downtime averaged over a period such as a year. Other possible compensation policies are based on the total number of failures experienced or the sum of squares of downtimes, etc. 
Such the relationship between physical time and monetary units connects traditional reliability analysis with business-oriented risk analysis.

\subsection{Related Work}

This research lays at the intersection of reliability analysis, risk management and machine learning.
Network resilience provisioning has been well studied~\cite{Schupke:Multilayer}. However, less emphasis is put to approaches aligned with the business perspective. 
Here, we assume risk-based quantification, where not only the probability of outages, but also their impact is directly taken into account. 
This way, we are able to characterize resilience in a 
more informative way than with the usage of classical resilience measures (e.g., reliability or availability functions). %
In the communications sector, risk has been associated with 
quantification of deviations from the desired quality levels, including 
security
~\cite{Teixeira:Secure}. %
From the mathematical viewpoint, the penalty is expressed on the basis of the risk theory dealing with extreme events. Generally, the most important aspect here is to estimate the whole distribution of the total penalty. Then, it is possible to quantify the penalty level with, for instance, the 95\textsuperscript{th} percentile, known as value-at-risk ($\mathit{VaR_{5\%}}$)~\cite{Wang:Value}. 
However, 
a more robust measure is suggested. It is called the conditional $\mathit{VaR}$ ($\mathit{CVaR}_{5\%}$). In this case, the average of all the penalties beyond the assumed percentile is calculated. An important advantage of $\mathit{CVaR}$ is its property of \textit{subadditivity}. It is helpful since with it we can assume as if the penalties for the individual connections were independent. Then, it is reasonable to sum the penalties for individual connections, since this sum is the upper bound for the total penalty in the network. Therefore, we can use the pessimistic approximation~\cite{RIGHI201514}. However, the common requirement for risk measures is the penalty distribution from which both measures can be derived. In the networking content, such distribution depends on the reliability parameters of network components.

In the modeling of up- and down-times, the typical approach is to assume that the failures arise due to a homogeneous Poisson process~\cite{Ahmad:Reliability}: 
the times between failures are exponentially distributed. 
It is statistically valid for many cases in communication networks~\cite{Gonzalez:Analysis}, but has appeared to be limited since other distributions for times between consecutive failures are also reported (e.g., Weibull distribution). %
The latter case considerably hinders the prediction of resilience parameters. %
Modeling of outage times is even more controversial. While the simplest approach also uses exponential times, these times in real networks appear to be log-normal~\cite{Garraghan:Analysis} or Pareto-like~\cite{Kuusela:On}. 
Combing both main fields it can be stated that machine learning algorithms have also been used in the context of risk management. A general framework for risk assessment using DL was proposed in~\cite{Paltrinieri:Learning} and validated in drive-off scenario involving an Oil \& Gas drilling rig. Authors provided comprehensive analysis and indicated several limitations of deep neural networks (DNNs) in terms of risk assessment. Moreover, DNNs were also successfully used for risk management in customs~\cite{Regmi:Risk} or financial market~\cite{Spyros:AIRMS}. 
The most prominent communication network models related to resilience are based on Markov theory. 
However, this type of modelling is not approachable and does not provide general and concise formulas for realistic cases, such as SBPP, where up and down-times do not follow simplistic assumptions of exponentiality. %
Due to the limitations of the modelling approaches, we decided to base on graph neural networks (GNNs)~\cite{ruiz2020graph}. The additional advantage of this approach is that it allows us to build a solution independent from a network topology and any particular 
routing scheme (by `routing' we mean any function mapping every SLA (connection) to the a pair of working/backup paths, i.e., sets of components (links))

GNNs have been already used for computer networks, for instance by Geyer et al.~\cite{Geyer2019}. %
The fundamental difference when comparing our work with this paper is that we consider routing paths as an input to the GNN-based algorithm, while~\cite{Geyer2019} aim at finding routing paths according to the specified policies. 
Additionally, Geyer et al. do not address the resilience aspect in the sense considered in our work. %
Namely, we are mainly focused on limitations in bandwidth of backup resources in SBPP.
The other papers of the same research group (e.g.,~\cite{Geyer:DeepTMA}) also differ with our work significantly: while they focus on control plane faults, we put emphasis on a topology-agnostic solution for the resilience of the data plane based on the business-oriented approach. %
Authors of~\cite{Sawada:Network} also focus on network performance by using GNN to extract features. They are used to calculate the routing that maximizes of bandwidth utilization. Resilience is also indirectly addressed, 
as the proposed solution is able to cope with router and link failures. 
Despite some similarities, our approach to resilience is more business-oriented (due to adoption of the risk approach) and not bounded to any particular method of generating working/backup paths.

\section{Methods}
\label{sec:methods}
The customers carry data on connections established in the network.
We assume that the network topology is represented by the following:
\begin{inparaenum}[(a)]
	\item Set $\mathcal{C} = \left\{ c_i \right\}_{ i= 1:n_c}$ of basic communication components $c_i$ (links) prone to failure. Only the link bandwidth can limit the shared protection capabilities. We assume realistic distributions of up- and down-times of links and treat them as the resilience attributes of these components. For the sake of this study, the routers are treated as fully reliable.
	\item Set $\mathcal{S} = \left\{\bm s_k \right\}_{k=1:n_s}$ of connections $\bm s_k$ between pairs of end-points (routers) in the network topology. 
	The routing for each SLA is defined as a pair of sets of components $\bm s_k = \left(s_p^k, s_b^k\right)$, where $s_p^k$ is the set of components in the working path and $s_b^k$ contains components in the backup path prepared for the $k$-th SLA. The connections are characterized by their demand volumes. These demands should be carried out with the help of resilient connections and this fact is reflected in service level agreements (SLAs).  
In essence, we identify a connection with its business-oriented description given by the SLA. 
\end{inparaenum}

In general, such properties (features) of both components and SLAs are denoted by vectors of data: $\mathbf x_{c_i}$ and $\mathbf x_{s_k}$, respectively. Vector $\mathbf x_{c_i}\in\mathbb{R}_{+}^4$ contains resilience parameters (parameters of up- and down-time distributions) and design parameters (i.e., backup bandwidth reserved in links) of individual component $c_i$. On the other hand, $\mathbf x_{s_k}\in\mathbb{R}_{+}^1$ contains 
SLA's parameters, in our case the demand volume for connection $\bm s_k$.
Both features $\mathbf x_{s_k}$, $\mathbf x_{c_k}$ and the routing $\mathcal{S}$ are jointly denoted as $\bm x=(\mathbf x_{s_k},\mathbf x_{c_k},\mathcal{S})$. %

\subsection{Approximate Penalty Distribution}

The ultimate goal in risk analysis of communication networks is the evaluation of the conditional distribution $P=\mathsf{P}(Y|\bm x)$ and in particular its quantiles to obtain $\mathit{VaR}$ or its derivatives. For simple protection schemes (e.g., dedicated protection) with the additional assumption of exponential up- and down-time distributions, $P$ can be obtained analytically~\cite{Rusek2016}.
According to the best authors knowledge, there are no analytical results for more realistic scenarios of shared protection procedures and non-Poisson downtimes.
Having said that, we point out that it is relatively simple to sample from $P$ by simulation.
In principle, one can estimate the risk measures with a Monte-Carlo method. However, the simulations take a lot of time to obtain reliable estimates due to the rare nature of the failures.
The method proposed in this paper is to approximate $P$ by a surrogate distribution $Q=\mathrm{nn}(\bm x)$, where $\mathrm{nn}$ is a neural network mapping topology properties to the family of parametric distributions.
The network consists of its own parameters (weights) $\theta$ learned from the simulations.

The training objective for $\mathrm{nn}$ used in this study is to minimize the Kullback–Leibler divergence $D_{\text{KL}}(P\parallel Q)$ between distributions.
In particular, for given network parameters and simulated penalty $\mathbf y\in\mathrm{R}^{n_s}$ we use Monte-Carlo approximation to the true KL divergence:
\begin{equation}
	D_{\text{KL}}(P\parallel Q) \approx\log\mathsf{P}_P(\mathbf y)-\log \mathsf{P}_Q(\mathbf y).
\end{equation}
Since $\log\mathsf{P}_P(\bm y)$  does not depend on $\theta$, we can ignore this term and use the negative log-likelihood function of the surrogate distribution as the loss function.
With the additional assumption of conditional independence of the SLAs, the loss function simplifies considerably to
\begin{equation}\label{eq:loss}
	\ell({\theta},\bm x, \mathbf{y}) = -\frac{1}{n_s}\sum_k \log \mathsf{P}_Q\left(y_k |\bm x,{\theta}\right),
\end{equation}

Due to its generalization potential, we applied the Student $t$-distributions as the parametric family. As used in many cases in statistical modeling (e.g., robust regression), we assume five degrees of freedom. This is a justified approach ensuring a proper heavy tail and is convenient for the training of neural networks (i.e., incurres a relatively simple likelihood function). Additionally, it ensures the existence of the mean value and variance. This way, we still have a bell-shaped distribution with an analytical PDF for efficient training of the $\mathrm{nn}$. Other candidate distributions involve normal and log-normal distributions. During the initial calculations, the normal distribution (suggested by the Central Limit Theorem) gave us similar results to $t$-distribution, upon inspection of a few simulations we observed that $t$-distribution matches the tail more accurately. On the other hand, log-normal distribution has a too heavy tail, and in this case, it is sufficiently accurate only for a smaller number of failures. The bell-shaped distribution is suitable for our simulation setup, as we always observe a few failures a year in a communication topology. In the case of an extremely resilient topology with the possibility of no-failures, a  zero-inflated log-normal distribution is recommended~\cite{mcdavid2019}, as it can model both the probability of no failure and the cost distribution given the failure has occurred.

The architecture of $\mathrm{nn}$ can be as simple as MLP, however, since there is no particular order of the SLAs nor the components, we additionally require model equivariance under permutations.
This makes a Graph Neural Network (GNN) a perfect candidate for $\mathrm{nn}$.

\subsection{GNN}

\begin{figure}
	\centering
	\includegraphics[width=\columnwidth]{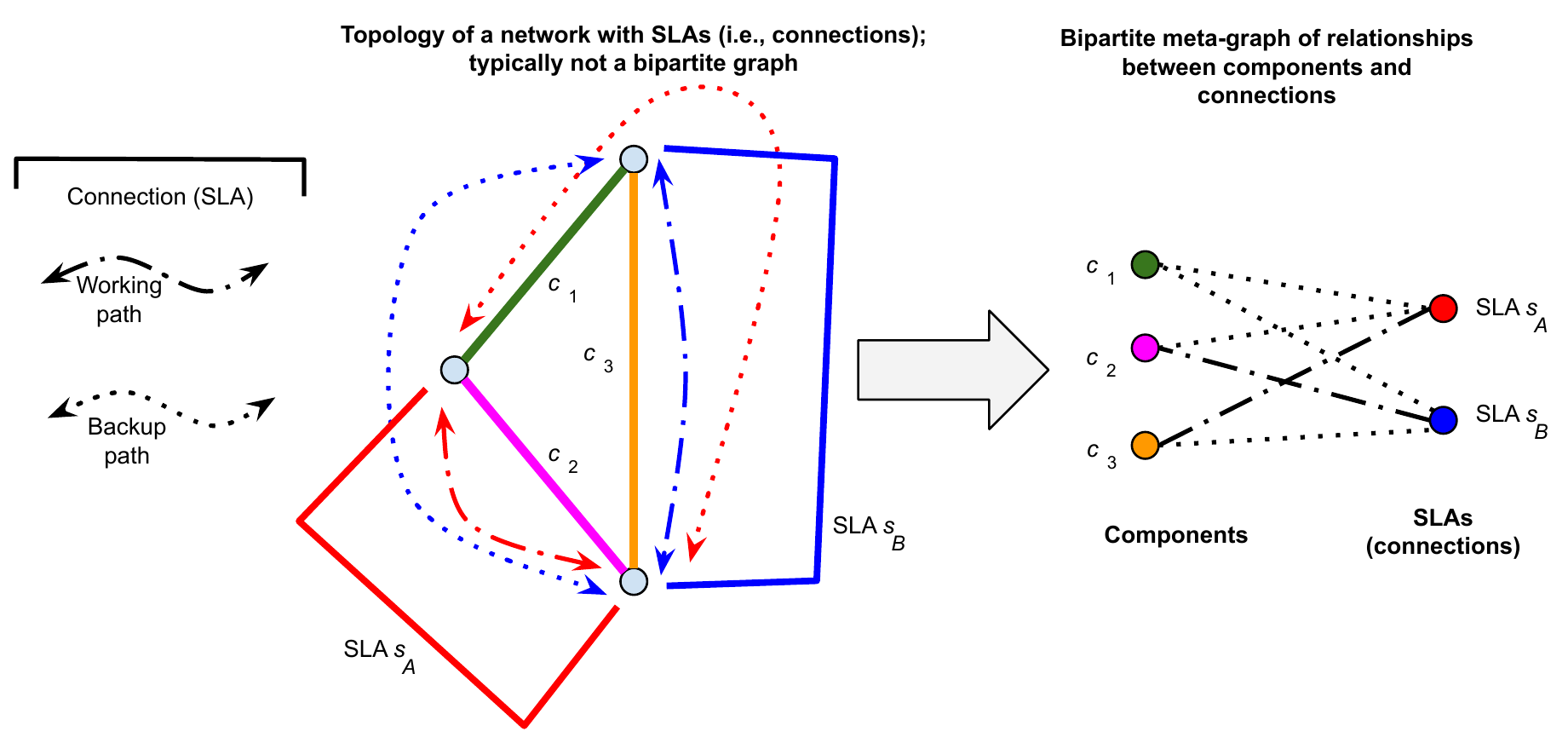}
	\caption{Transformation of the given topology onto the associated bipartite metagraph for GNN}
	\label{fig:prediction-bipartite}
\end{figure}

The association between the set of components and the set of SLAs can be represented as a \textit{bipartite metagraph} (for simple illustration, see Fig.~\ref{fig:prediction-bipartite}). Both components and SLAs can be represented as nodes\footnote{To avoid misunderstanding, the vertices of the bipartite metagraph are denoted as `nodes', while the vertices of the studied communication network topology are called `routers'.} of the association graph, with edges connecting an SLA and a component if and only if the latter is used in the path for the given SLA. There are two kinds of edges in the metagraph: one representing a working path (i.e., that a component is used in SLA's working path), and the other for the backup path.

Our main idea is to run a heterogeneous GNN named RiskNet on the presented bipartite metagraph to get the surrogate penalty distribution. We base on a version of GNNs known as \emph{message-passing neural networks}.
Such a network embeds the knowledge of internal relationships in the vectors associated with the component and SLA-related nodes, denoted 
as $\bm h_c^{t}$ (for component $c$) and $\bm h_s^{t}$ (for SLA $s$), respectively.
These vectors, known as internal (hidden) states, are found as one of the results of our algorithm operation.
The forward pass begins with zero-padded components and SLA feature vectors, followed by an iterative exchange of messages and state updates. In particular, the result of embedding every edge in the metagraph gives vectors $\tilde{\mathbf m}_{r,u\to z}^t$, where $t$ represents the iteration; $r \in\{p,b\}$ represents the message related either to working ($p$) or backup ($b$) path, and $u,z\in\{c,s\}$ represent in which direction the message is passed ($s\to c$ denotes SLA to component message, while $c\to s$ denote the message in the opposite way).
The message is calculated as an output of a neural network represented as $\tilde{\mathbf m}_{r,u\to z}^{t+1}=M_{r,u\to z}^{t}(\bm h_u,\bm h_z)$ (the abovementioned notation related to $\tilde{\mathbf m}_{r,u\to z}^t$ is again valid). 
These neural networks are called \textit{message functions}. 
We can see that the working and backup paths have different message functions, and they can deal with two directions. Therefore, we have four message functions used 
($M_{p,s\to c}^{t}$, $M_{b,s\to c}^{t}$, $M_{p,c\to s}^{t}$, $M_{b,c\to s}^{t}$). These functions can be also different in various iterations (that is, why we also use the superscript $t$ for them).
To calculate the change of a hidden state, 
we use neural networks denoted as $U_c^t$ and $U_s^t$. These \textit{update functions} encode the combined incoming information into the hidden state:
$\mathbf h_v^{t+1} = U_v^t\left(\mathbf h_v^t,\mathbf m_v^{t+1}\right)$,where $v\in\{s,c\}$ and $\mathbf m_v^{t+1}$ is the sum of the embedding vectors for all adjacent edges in the metagraph.

The whole process is typically convergent after a few ($T$) iterations. This parameter controls the range of interactions between SLAs, e.g., for $T=1$ each SLA receives messages only from its components and the model is basically a DeepSet~\cite{NIPS2017_deepset}. 
Higher $T$ allows for information to be exchanged between SLAs trough components. 
The predicted penalty related to an SLA is found using a small \textit{readout} neural network (represented as $F$) applied to the final hidden state of the SLA. 
Its output represents the parameters of the penalty distribution for this particular SLA. 
In terms of particular neural architectures we used \textit{affine functions} for message propagation ($M$) and \textit{GRU units} for the update ($U$). 
These are the proven units used in many GNN architectures. They are selected as a trade-off between simplicity, performance, and flexibility of the model. 
Similarly to the previously proposed models, the weights of message and update functions are reused for subsequent message-passing iterations. 
The readout function proposed in this paper is a multilayer perceptron with the $\mathit{SeLU}$ activation function. The mapping from the raw readout output to Student $t$-distribution's location parameter is the identity; however, the scale must be constrained to positive numbers, so we use the softplus function.

\subsection{Simulation}

All experiments reported in this paper are supported by the discrete event risk simulator previously used and verified by comparison with the theoretical results in~\cite{Rusek2016} in series of unit tests. The software is written in C++. 
The simulator is treated as the source of ground truth and it is used for the creation of training datasets for neural network (GNN). 
A network topology is the starting point for building the configuration. In the experiments, the topologies under study are either the ones based on the random  Barab\'{a}si--Albert model (with the output power distribution of node degrees) or the existing topologies retrieved from the SNDLib library (\url{http://sndlib.zib.de}).
According to the random model, every new router is attached to at least two existing ones.
This method always makes it possible to construct the working and backup path for any connection between a pair of routers. %
Afterwards, for a selected topology, first we generate the working and backup path for every connection (SLA) between all routers. 
For the training dataset creation, we do not use a typical approach of looking for 2-shortest candidate paths. Instead, we choose the router-disjoint path randomization from the set of all disjoint paths found for all connections. Namely, we set the probability of drawing a given path 
as decreasing in the function of a path's length (e.g., a number of components on the path). We select a parameter $\xi$ and then draw paths with probability $p_k\sim e^{-\xi |s_k|}$. This value is normalized. Then, for small values of $\xi$, the distribution is flat and long paths have a high probability to be selected, while for large $\xi$, we have mainly the shortest paths, as $p_k$ drops quickly with the length.
For the training phase, we use $\xi=0.1$. This approach allows us to explore the configuration space and gives the GNN a highly diverged set of training samples. %
Demand volume of a connection is proportional to the product of the sizes of source and sink routers for the connection. The size of a router depends on its degree $d$ and is uniformly distributed in the interval $10\times(d\pm1)$. %
At the end, the resilience parameters of different components are generated. 
Here, we assume Poissonian failures (i.e., exponential up-times) and Pareto-distributed down-times. Both are parametrized according to the estimates reported in~\cite{Kuusela:On}. Link lengths are obtained from the scaled spring layout of the network topology. %
The output of the simulation is the total penalty for each SLA in every simulated year of operation. This value is used as a target (label value $y$) in the process of GNN training.

\section{Numerical Results} \label{sec: evaluation-of-the-accuracy}

To simplify, we can say that we use a type of supervised learning (heteroscedastic regression), since we are able to provide the real values (ground truth) of the penalties and confront them with the predicted values to train the model.
The real values are provided by simulations for specific configurations. On the other hand, simulations take a lot of time, that is why we can afford for them only in the training phase. During operation of the whole system, a very fast GNN model predicts the values for given working/protection path settings. 

\subsection{Training and hyperparameters}

During the training phase, first we select different network topologies with randomly generated parameters (training samples). 
They fed the GNN and a discrete-event simulator and the result is stored for offline training. 
Second, the message-passing is run in GNN to obtain the convergence 
and provide prediction of the total penalties $\hat y$. 
The predicted penalties are confronted with $y$, i.e., the ground truth (real penalty levels) provided by the simulator. On this basis, the learning error (loss) is calculated 
and the classical backpropagation algorithm is used to update the internal weights of the neural networks forming GNN. 
As concerns the inference, 
the output from the prediction model follows the same path as in the case of the training. 
The only exception to when one wants to use dropout to estimate the uncertainty of the prediction. Then, the output is equal to the average of multiple stochastic forward passes. 

In the spirit of the modern deep learning approach, we used the raw simulator parameters as the input to RiskNet and let the model learn a meaningful internal representation. 
The SLA feature contains the demand volume only. 
The components' feature vector has four dimensions 
(failure intensity, $\alpha$ and $\beta$ parameters of the Pareto downtime distribution, and the capacity reserved for protection). The only transformation applied to the data is the z-score normalization, as it improves the training process. We do not consider routing as a feature, but rather as metadata. %

Training of a deep neural network typically requires hundreds of thousands of samples. Learning from simulations makes it easier to obtain samples; however, we are still limited by the training and simulation time. Our training set is constructed from a simulation of 1000 random network topologies (generated according to Barab\'{a}si--Albert model) 
with a number of routers uniformly distributed in the range $[10,40]$. Every topology was simulated for up to 1000 years of operation.
Since for some of the largest networks not all simulations finished in the assumed time constraint, we ended up with around 829\,000 training examples. Using the same procedure, we generated additional 20\,000 test samples to spot signs of overfitting. 
With this training set, we tested multiple configurations of the RiskNet mostly differentiated by the hyperparameter values. On this basis and according to our prior knowledge about GNNs, we chose the final configuration. Both hidden vectors have 32 dimensions. The message has 64 dimensions. 
The kernel of the affine message function is regularized with the coefficient $0.01$, and the bias is not regularized. 
The message passing loop is iterated six times. 
Finally, the readout function has three hidden layers of sizes $(64,64,32)$ interleaved with two dropout layers with dropout rates 0.2 and 0.1, respectively. 
The model characterized above was trained for 54 epochs of 12\,900 iterations of the Adam optimizer on a batch of 64 topologies. %
The learning rate was set to 0.0001 for the first 20 epochs. Then it was decayed by 0.99 per epoch. The whole training took 11\,hours\,40\,min with $T=6$. 

The model was implemented entirely in TensorFlow.
The hardware supporting calculations embraces 36 cores Intel(R) Xeon(R) Gold 5220 CPU @ 2.20GHz, while the graphics processor unit used is GPU Tesla V100 SXM2. As concerns the effective speed of calculations, we emphasise that they are extremely fast. %
For instance, the average calculation 
time for the janos-us network equals 4$\pm$0.2\,ms for a single evaluation of the penalty by GNN (11\,000 evaluations per minute). On the other hand, a single simulation of 600\,years of network operation takes 49$\pm$2\,sec. with 36 cores of CPU. This way, we obtain 12\,000 times speed improvement of calculation. %

\subsection{Evaluation}
The final model was evaluated with the third synthetic dataset as well as with the majority of topologies retrieved from SNDLib.
To show the benefits of RiskNet, we compare the results to the baseline model. Here, the baseline is a marginal distribution of all penalties, i.e., the distribution of penalties in all experiments without any distinction based on $\bm x$. Since the training data was normalized, the baseline distribution is a standard Student $t$-distribution with five  degrees of freedom.
GNN can improve prediction by using information from the features $\bm x$ as summarized in Table~\ref{tab:nnval}. 

Due to the fact that GNN estimates the whole distribution, the common metrics, such as mean squared error or mean percentage error are no longer meaningful.
The loss must be expressed with negative log-likelihood.
Therefore, the negative or positive value is not easy to interpret. 
However, the smaller value, the better and one can note a significant improvement over the baseline provided by our approach.
Furthermore, we can see that the results of the test evaluations are close. 
This proves the generalization capabilities of the model. 
The effect is further confirmed by results on the real topologies, where GNN provides average scores at the level of -0.88 vs. the baseline result of 1.62. 

\begin{table}
	\centering
	\caption{GNN evaluation loss}
	\label{tab:nnval}
\footnotesize
	\begin{tabular}{lrrr}
		\toprule
		\bfseries Topology &  \bfseries GNN $T=6$ &  \bfseries Baseline & \bfseries GNN \bfseries $T=1$\\
		\midrule
		train          &     -1.37 &            -- & -1.34\\
		test           &     -1.43 &            -- & -1.38\\
		validation     &     -1.42 &           1.10 & -1.39\\
		Average SNDLib &     -0.88 &           1.62 & -0.74\\
		dfn-bwin       &      0.73 &           4.09 & 1.19\\
		abilene        &     -1.67 &           0.87 & -1.62\\
		nobel-germany  &     -1.37 &           0.88 &-1.32\\
		cost266        &     -1.32 &           1.00& -1.25 \\
		geant          &     -1.38 &           1.01 &-1.33\\
		nobel-eu       &     -1.32 &           0.97 & -1.20\\
		janos-us       &     -1.14 &           1.11 & -1.04\\
		\bottomrule
	\end{tabular}
\end{table}

The loss obtained for RiskNet is much lower compared to baseline. 
This indicates that the model indeed learns the information from a network topology. 
We can make this statement more precise in the context of information theory. 
The difference between log-likelihoods is a measure of information the model has learned.
By changing the base of the logarithm to 2, we can express this information in bits.
In particular, for the validation set, the RiskNet system reduces the description length on average by 3.6 bits per path (in comparison to the baseline).
For reference, the entropy of the baseline distribution is 1.6 bits.

\begin{figure}
	\centering
	\includegraphics[width=\columnwidth]{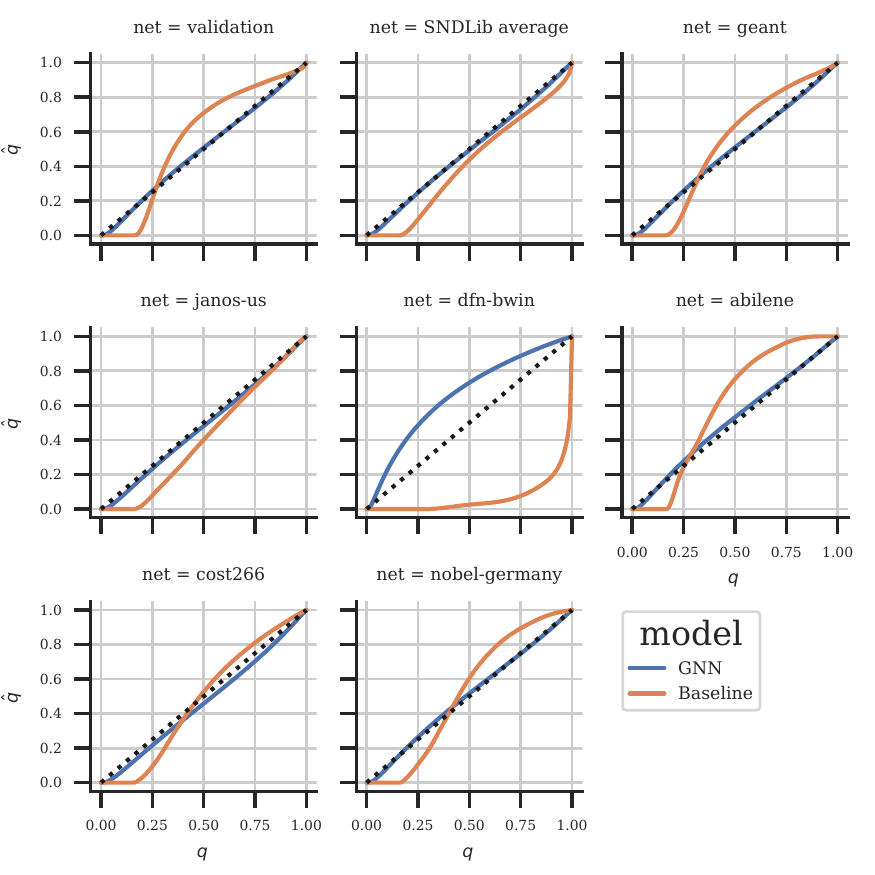}
	\caption{Probability plots for topologies simulated for 1000 years of operation }
	\label{fig:pp}
\end{figure}

Despite the fact that the negative log-likelihood applied as a loss function is a theoretically justified measure used for parameter estimation, it is difficult to state how accurate the fit is by reasoning on the basis of a single value. Therefore, to provide a more intuitive measure, we propose to use probability plots (pp-plots, see Fig.~\ref{fig:pp}) 
produced according to the following procedure. 
For every network configuration $\bm x$, RiskNet predicts the whole distribution of penalties $Q$. The ground truth $\bm y$ is obtained from the simulation. Given some probability value $q$, we construct a Bernoulli random variable $\mathsf{1}_{\bm y<\bm y_q}$, where $y_q$ is a $q$-quantile of $Q$. 
If $\bm y$ were sampled from $Q$, the probability of this Bernoulli distributed random variable would be equal to $q$. Since the predicted distribution is not the exact sampling distribution $P$, the estimated probability $\hat q$ will be different. The closer it is to $q$, the better approximation of the distribution we obtain. 
For the Bernoulli distribution, the unbiased estimator of the probability is just the average value, so we use $\hat{q} = \overline{\mathsf{1}_{\bm y<\bm y_q}}$. The pp-plot is constructed as a line plot of $\hat{q}$ vs. $q$. The diagonal line is added for the reference. Any deviation from this line indicates a mismatch in distribution. 
We can observe that the distribution produced by RiskNet is much closer to the diagonal line compared to the baseline. What is even more important is the fact that the deviation from the diagonal is small for all probabilities. This tells us that RiskNet correctly predicts multiple quantiles of the distribution. This is a highly required feature, as it allows us to use the same model for estimation of risk at different levels (namely, various percentiles $p$ of $\mathit{CVaR_{(1-p)\%}}$). This property is practically useful when we would like to estimate various risk levels.

The only case where RiskNet distribution significantly deviates from the empirical samples is for the \emph{dfn-bwin} network. 
However, this network is dissimilar to distribution samples. 
The network topology is almost the full graph in contrast to small-world topologies produced by the Barab\'{a}si--Albert model.
The fact that the baseline is also much less accurate for this network supports our claims even further.
One must remember that RiskNet is a statistical model, and despite its generalization capabilities, there are some edge cases where the model can no longer be considered as a good approximation.
On the other hand, for some cases a simpler model can be acceptable.

In the experiments, we considered simpler models with weaker interactions between SLAs (as measured by $T$). In particular, setting $T=1$ yields a surprisingly accurate model\footnote{The information difference between models is 0.04 bit per SLA.} (see Table~\ref{tab:nnval}) without any interactions and whose test loss quickly begins to grow during the training. The similar behaviour was observed in other weekly interacting models --- all of them suffered from overfitting. We conclude that a high value of $T$ acts as a regularizer for the model.  We explain this by the fact that networks in the simulation were highly reliable and most of the contributions to the penalty were due to a single failure only. Having said that, we emphasize that in general SLAs must exchange information since, by definition, they do interact in the case of shared protection.

\section{Conclusions} \label{sec:conclusions}

In this letter, we show that a GNN can parametrize the t-Student distribution to approximate the penalty cost distribution due to component failures in a network. Even though this work is derived and motivated by our experience in the field of communication and computer networks, it generalizes well beyond this area. Similar concepts of network or unreliable components arise in logistics and other areas of business importance. Since the penalty under consideration is based on downtimes, we expect this work to be extendable to the downtime related quantities.

Our approach allows us to
\begin{inparaenum}[(a)]
	\item omit very complex, time-consuming, and ineffective modeling of resilience parameters for shared protection;
	\item improve practically useful prediction quality of business-oriented risk parameters by using an ML-based module, providing very good results in comparison to the baseline case;
	\item replace time-consuming simulations, being an intuitive alternative to our proposal, by a very fast prediction method used --- to can be applied by a network designer during connection optimization.
\end{inparaenum}






\end{document}